\documentclass[aps,prx,twocolumn,superscriptaddress,showpacs,floats,a4paper,nobibnotes]{revtex4-1}
\usepackage{newfloat}
\usepackage{latexsym}
\usepackage{dcolumn}
\usepackage{graphicx}
\usepackage{amssymb}
\usepackage{amsmath}
\usepackage{amsfonts}
\usepackage{mathrsfs}

\usepackage{float}
\usepackage{hyperref}
\hypersetup{colorlinks,linkcolor=blue,citecolor=blue,urlcolor=blue}
\usepackage[left=18mm,right=18mm,top=25mm,bottom=25mm]{geometry}
\usepackage{bm}
\usepackage{mhchem}
\usepackage[section]{placeins}
\usepackage{color}

\begin{document}

\title{Transport and magnetotransport in 3D Weyl Semimetals}

\author{Navneeth Ramakrishnan}
\affiliation{Department of Physics and Center for Advanced 2D Materials, National University of Singapore, 117551, Singapore}

\author{Mirco Milletari}
\email{phymirc@nus.edu.sg}
\affiliation{Department of Physics and Center for Advanced 2D Materials, National University of Singapore, 117551, Singapore}

\author{Shaffique Adam}
\email{shaffique.adam@yale-nus.edu.sg}
\affiliation{Department of Physics and Center for Advanced 2D Materials, National University of Singapore, 117551, Singapore}
\affiliation{Yale-NUS College,  16 College Avenue West, 138527, Singapore}

\date{\today}

\begin{abstract}
We theoretically investigate  the transport and magnetotransport properties of three-dimensional Weyl semimetals.  Using the RPA-Boltzmann transport scattering theory for electrons scattering off randomly distributed charged impurities, together with an effective medium theory to average over the resulting spatially inhomogeneous carrier density, we smoothly connect our results for the minimum conductivity near the Weyl point with known results for the conductivity at high carrier density. In the presence of a non-quantizing magnetic field, we predict that for both high and low carrier densities, Weyl semimetals show a transition from quadratic magnetoresistance (MR) at low magnetic fields to linear MR at high magnetic fields, and that the magnitude of the $MR \gtrsim 10$ for realistic parameters.  Our results are in quantitative agreement with recent unexpected experimental observations on the mixed-chalcogenide compound TlBiSSe.  
\end{abstract}
\maketitle

\section{Introduction} Electronic band structures that have protected gapless points -- where the conductance and valence bands are guaranteed to meet -- have been of significant theoretical and experimental interest in recent years.  The two dimensional manifestation of such band structures have been extensively studied in graphene~\cite{sarma2011electronic}, where the gapless nature is protected by sublattice symmetry~\cite{neto2009electronic}, and in 3D topological insulators~\cite{HasanKaneRMP}, where the crossing point is protected by topology~\cite{KaneMelePRL, kane2005z}.  More recently, attention has focused on the three dimensional analogues of these compounds, called Weyl semimetals~\cite{AshwinTheoryPaper, wang2012dirac, burkov2011weyl}.  Compounds such as Cd$_{3}$As$_{2}$~\cite{borisenko2013experimental, ali2014crystal, ongpaper} TaAs~\cite{xu2015discovery} and TlBiSSe~\cite{xu2011topological, singh2012topological} have been shown to have Weyl points in their band structure (see Ref.~\cite{gibson20143d} for a recent review on the various candidate materials for semimetals with 3D relativistic electronic dispersions). 

Theoretical efforts toward characterizing the electronic properties of Weyl semimetals are in the nascent stage and include the scattering properties of different impurity potentials~\cite{ominato2014quantum}, localization and delocalization~\cite{sbierski2014quantum, lu2014tendency}, thermoelectric properties~\cite{lundgren2014thermoelectric}, screening~\cite{zhang2013,roy2014diffusive} and temperature dependence~\cite{hwang2014carrier}, the influence of the chiral anomaly~\cite{burkov2014chiral}, diffusive transport~\cite{biswas2014diffusive} and the effects of electron-electron interactions~\citep{witczak2014interacting}.  Inspired by unexpected observations in recent transport~\cite{ongpaper, novak2014large} and scanning probe~\cite{jeon2014landau} experiments, we study theoretically the transport and magnetotransport properties of 3D Weyl semimetals in the presence of randomly distributed Coulomb impurities. The effect of the charged impurities is twofold: they provide a momentum relaxation mechanism and they act as dopants for the local carrier density. At low carrier density, the former mechanism introduces macroscopic inhomogeneities in the carrier density profile, giving rise to positively or negatively charged puddles.   

In this work we use a random phase approximation (RPA) method to evaluate the effective screened potential of the Coulomb impurity that enters various disordered averaged quantities. We show that the RPA is a much better approximation of the commonly used Thomas Fermi approximation due to the nature of the vacuum screening structure of Dirac materials. In the homogeneous regime (far from the Weyl point in momentum space), weak impurity scattering is considered at the Born level to obtain the Drude conductivity. In the inhomogeneous regime (close to the Weyl point in momentum space), we find that it is important to consider structure in the disorder distribution when performing the disorder average. We then use an Effective Medium Theory (EMT) to average over the  inhomogeneous carrier density distribution. This formalism has been remarkably successful in providing a quantitative understanding of the transport properties close to the Dirac point in graphene~\cite{sarma2011electronic}, the 2D cousin of these Weyl semimetals. Our results allow us to make both quantitative and qualitative comparisons with experiment and yield many qualitative insights into the behavior of Weyl semimetals under various experimental conditions. 

This paper is organized as follows: In Section~\ref{Transport at High Carrier Density}, we discuss the Drude conductivity using both the Thomas-Fermi and RPA screening approximations, far away from the Weyl point and also discuss why the RPA is required (unlike in the case of graphene). In Section~\ref{Transport at Low Carrier Density}, we discuss the effects of impurity correlations and induced charge carriers, that play a role near the Weyl point. Finally, we look at experimental results from Ref.~\cite{novak2014large} and compare them with our theoretical models in Section~\ref{Comparison with experiments}. The experimental transport data is found to be consistent with two possible theoretical regimes and we note in Section.~\ref{Magnetoresistance} that magnetotransport provides a simple and experimentally accessible mechanism to distinguish between the two possibilities. 

\section{Transport at High Carrier Density}
\label{Transport at High Carrier Density}
The Hamiltonian for a Weyl semimetal is given by

\begin{align} \label{Ham}
H&=\pm \, \hbar \,\imath \, v_{\it F} \, \boldsymbol{\sigma} \cdot \boldsymbol{\partial}_{\mathbf{r}} - \xi
 + V(\mathbf{r}) \\ \nonumber
V(\mathbf{r})&=\sum\limits_{j=1}^{N_{\rm imp}} U(\mathbf{r}-\mathbf{R}_{\rm j}) 
\end{align}
where $\boldsymbol{\sigma}$ is a vector of Pauli matrices, $\xi$ is the chemical potential, $v_{\it F}$ is the Fermi velocity and $\pm$ accounts for the two chiralities. The density of states is $\nu(E)= g |E|^2/2 \pi^2 v_{\it F}^3$, where $g$ is the the degeneracy (here $g=4$ due to spin and the presence of two cones). In Eq.~\eqref{Ham}, $U(\mathbf{r}-\mathbf{R}_{\rm j})$ is the total screened potential seen by an electron at position $\mathbf{r}$ due to charged impurities at positions $\mathbf{R}_j$. In this work, we consider Coulomb impurities with momentum space screened potential
\begin{equation}
U(\mathbf{q}) = \frac{4\pi e^2}{\epsilon(\mathbf{q}) \mathbf{q}^2},
\end{equation} 
where $e$ is the electronic charge and $\epsilon(\mathbf{q})$ is the dielectric function. Here $\mathbf{k}$ and $\mathbf{k'}$ are the incoming and outgoing momenta of the scattered electron and $\mathbf{q}=\mathbf{k}-\mathbf{k'}$ is the transferred momentum. Note that the Coulomb potential suppresses large momentum scattering connecting the two Weyls cones. For this reason, in the following we will work with one cone and consider the contribution of the second one in the degeneracy factor. 
For a given concentration of impurities, $n_{\rm imp}$, the ensemble averaged transport scattering  time within the Born approximation is given by
\begin{equation} \label{transtau}
\frac{\hbar}{\tau_{tr}} = 2\pi \, n_{\rm imp} \int \frac{d^3\mathbf{k'}}{(2\pi)^3} \, U(|\mathbf{k}-\mathbf{k'}|)^2 \frac{1-\cos^2\theta}{2}\delta(E_{\rm k}-E_{\rm k'}).
\end{equation}
To make connection with existing results in the literature~\cite{hwang2014carrier, skinner2014coulomb} we first consider the simpler case of evaluating Eq.~\eqref{transtau} in the Thomas-Fermi (TF) approximation, where the scattering potential can be taken as $U(q) = 4\pi e^2/(\kappa(q^2+q_{\rm s}^{2}))$ with a transferred momentum, $q=|\mathbf{k}-\mathbf{k'}|$. Here $\kappa$ is the dielectric constant of the material and $q_{\rm s} = \sqrt{4\pi e^2 \nu(E_{\it F})/\kappa}$ is the inverse of the Thomas-Fermi screening length.  Introducing the effective fine structure constant $\alpha=e^2/\hbar v_{\it F} \kappa$ (where $\alpha = 0.07$ for Cd$_{3}$As$_{2}$ and $\alpha = 0.68$ for TlBiSSe), we have~\cite{burkov2011topological, hwang2014carrier} 
\begin{equation} \label{condTF}
\sigma_{\rm TF} =\frac{e^2 v_{\it F}^{2}\tau}{3} \nu(E_{\it F}) 
=\frac{e^2}{h}\frac{g}{12\pi^2}\frac{k_{\it F}^4}{n_{\rm imp} \, \alpha^2}\frac{1}{H(\sqrt{\frac{g\alpha}{2\pi}})},
\end{equation}
where $H(z) = (z^2+ 1/2)\log(1+z^{-2})-1$.
%
% \begin{equation} \label{transtauTF}
% \frac{1}{\tau}= 4\pi \, n_{\rm imp} \, \alpha^2 \, \frac{v_{\it f}}{k_{\rm f}^2} \, I\left(\sqrt{\frac{g\alpha}% {2\pi}}\right),
% \end{equation}
%
In both non-chiral two dimensional electron gases and graphene, the accidental coincidence at zero temperature of the polarization function for $q \leq 2 k_{\it F}$ and the density of states implies that the Thomas-Fermi approximation gives identical results to the RPA.  However, we should emphasize that this is no longer true at finite temperature or in other materials like bilayer graphene~\citep{sarma2011electronic}.  As we discuss below, for Weyl semimetals the Thomas-Fermi and RPA give qualitatively different results, and in what follows, we use the more correct RPA approximation with $q_{\rm s}(q) =  [\frac{e^2}{\kappa}D(E) \, \tilde{\Pi}(q/2 k_{\it F})]^{1/2}$ . Here, $\tilde{\Pi}(x)$ is the ratio of the RPA polarization function and the density of states, and is given by the sum of two components: a vacuum part  $\tilde{\Pi}_V(x)$ and a finite density part $\tilde{\Pi}_M(x)$~\citep{zhang2013}

\begin{align} \nonumber
\tilde{\Pi}_M &= \frac{2}{3} \left[ 1+\frac{1}{4x}(1-3x^2)\log \left| \frac{1+x}{1-x}\right| - \frac{x^2}{2}\log\left|\frac{1-x^2}{x^2}\right| \right] \\ 
\tilde{\Pi}_V &= \frac{2x^2}{3}\log\left|\frac{\delta}{x}\right|.
 \label{pol} 
\end{align}

\begin{figure}[!t]
	\centering
		\includegraphics[angle=0,width=0.5\textwidth]{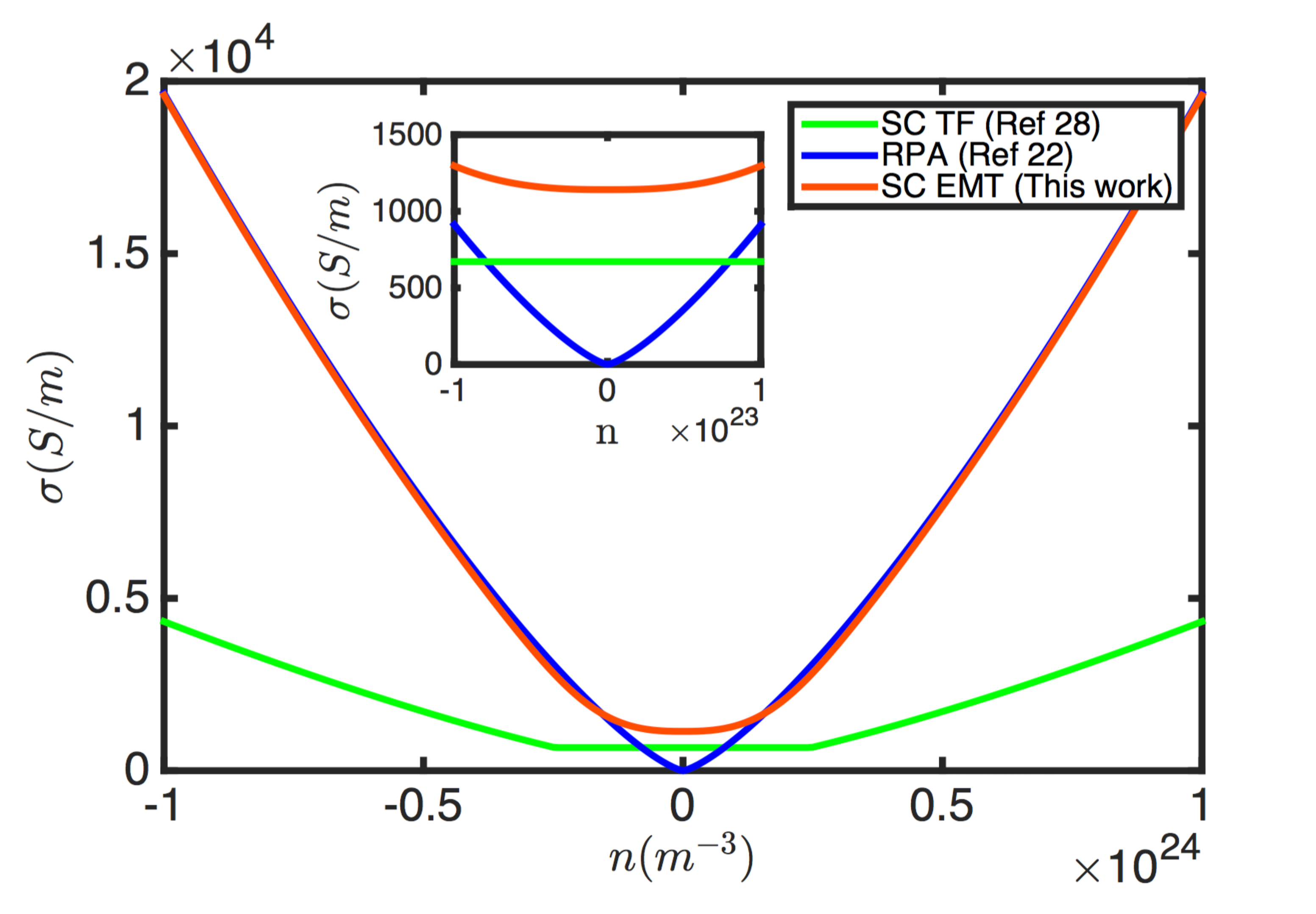}
	\caption{ {\bf Conductivity as a function of carrier density.} RPA-Drude theory (Ref.~\citep{hwang2014carrier}, blue curve), self-consistent Thomas-€"Fermi (Ref.~\citep{skinner2014coulomb} , green curve) and the EMT approach (assuming uncorrelated impurities in both regimes) discussed in the main text (red curve).  The EMT predicts a minimum conductivity close to the Weyl point and reproduces the Drude one $\sigma \sim n^{4/3}/n_{\rm imp}$ far away from the Weyl point. In order to compare with the results of Ref.~\citep{hwang2014carrier}, we follow their parameters for this figure and use $\alpha=1.2$, $g=2$ and $n_{\rm imp} = 10^{24}m^{-3}$ and $\tilde{\Pi}_{V}=0$.  (Inset) Close-up of the Weyl point.}
	\label{fig:sig_vs_n}
\end{figure}

We remark that the Thomas Fermi calculation is merely the linearized result of the RPA which is valid for $q\ll k_F$~\citep{Ashcroft}. Since we consider transport both near and away from the Weyl point, the TF approximation is insufficient. Moreover, the TF approximation neglects the vacuum term of the polarization function, which in our case is the dominant term for $q\gg k_{F}$. As seen in Figure~\ref{fig:sig_vs_n}, the large quantitative difference between the two approximations implies that one must perform the full numerical RPA calculation for accurate comparison with experiments. Finally, we would like to remark that unlike the case of one and two dimensions, for Weyl fermions in $d=3$ the vacuum polarization function is divergent and needs an ultraviolet momentum cutoff $\Delta$, where in Eq~{\eqref{pol}}, $\delta=\Delta/2k_{\it F}$.  In principle, the transport coefficients calculated within the RPA approximation could depend on the choice of the cutoff~\cite{cutoff}, although in practice we have verified that such dependence is weak for a realistic parameter range. Our results for the high density transport in the absence of a magnetic field are shown in Fig.~\ref{fig:sig_vs_n}.  We find that far away from the Weyl point, $\sigma\sim n^{4/3}/n_{\rm imp}$, where $n$ is the carrier density. In Fig.~\ref{fig:sig_vs_n}, we show that this result is in  agreement with calculations recently reported in Ref.~\cite{hwang2014carrier} for the homogeneous regime.

\section{Transport at Low Carrier Density} 
\label{Transport at Low Carrier Density}
\subsection{Correlations in Impurity Positions}
In the regime near the Weyl point, we consider two different effects. First, the conductivity may be modified from the homogenous case due to the presence of correlations in the impurity positions. In order to understand the origin of these correlations, one notes that the effective ``size" of the impurity potential is approximately $r_0 \simeq \lambda \simeq n^{-1/3}$, where $\lambda$ is the screening length. This should be compared with the average distance between the impurities, given by $L \simeq n_{imp}^{-1/3}$. At high carrier density, screening is more effective and the screening length is smaller, corresponding to $r_0 \ll L$ (or $n \gg n_{imp}$). In this regime, impurities are well separated and can be essentially considered as point-like, meaning that disorder is completely random. In the inhomogeneous regime however, we find that the electron density is comparable to the impurity density and therefore $r_0 \sim L$. In this case, disorder cannot be considered as completely structureless and correlations between impurity positions need to be included~\cite{ziman1979models, li2012effect}. Since the pair correlation function $g(\mathbf{R}_i, \mathbf{R}_j) \neq g(\mathbf{R}_i) g(\mathbf{R}_j)$, the self energy term proportional to $n_{imp}^2$ cannot simply be renormalized away. Finally, assuming that the impurities are homogeneously and isotropically distributed, the pair correlation function reduces to the radial correlation function $g(R)$. The effect of correlations can be taken into account by the modified correlator of the Gaussian random fields (see Appendix~\ref{sec:corr})
\begin{align}
\label{eq:rel} 
\langle V(\mathbf{r}) V(\mathbf{r}') \rangle_d &= n_{imp} \int d^3q \, e^{\imath \, \mathbf{q} (\mathbf{r}-\mathbf{r}')} \, U^2(q) \, S(q)  \\ \nonumber
S(q) &= 1+ n_{imp} \, \int d^3 R \, \left\{ g(R)-1 \right\} e^{-\imath \, \mathbf{R} \cdot \mathbf{q} },
\end{align}
where $\langle ... \rangle_d$ stands for disorder average and we have introduced the structure factor $S(q)$. Following Ref.~\cite{li2012effect}, we take the radial distribution function as
 \begin{equation} \label{eq:gr}
g(R)= \left\{\begin{array}{c}0 \quad R<r_0 \\ 1  \quad R>r_0 \end{array}\right . 
\end{equation}
When computing transport properties, the effect of Eq.~\eqref{eq:rel} is to replace $ U^2(q) \to U^2(q) \, S(q)$ in Eq.~\eqref{transtau}. 

\subsection{Effective Medium Theory} \label{sec:emt}

\begin{figure}[!htb]
	\centering
		\includegraphics[angle=0,width=0.50\textwidth]{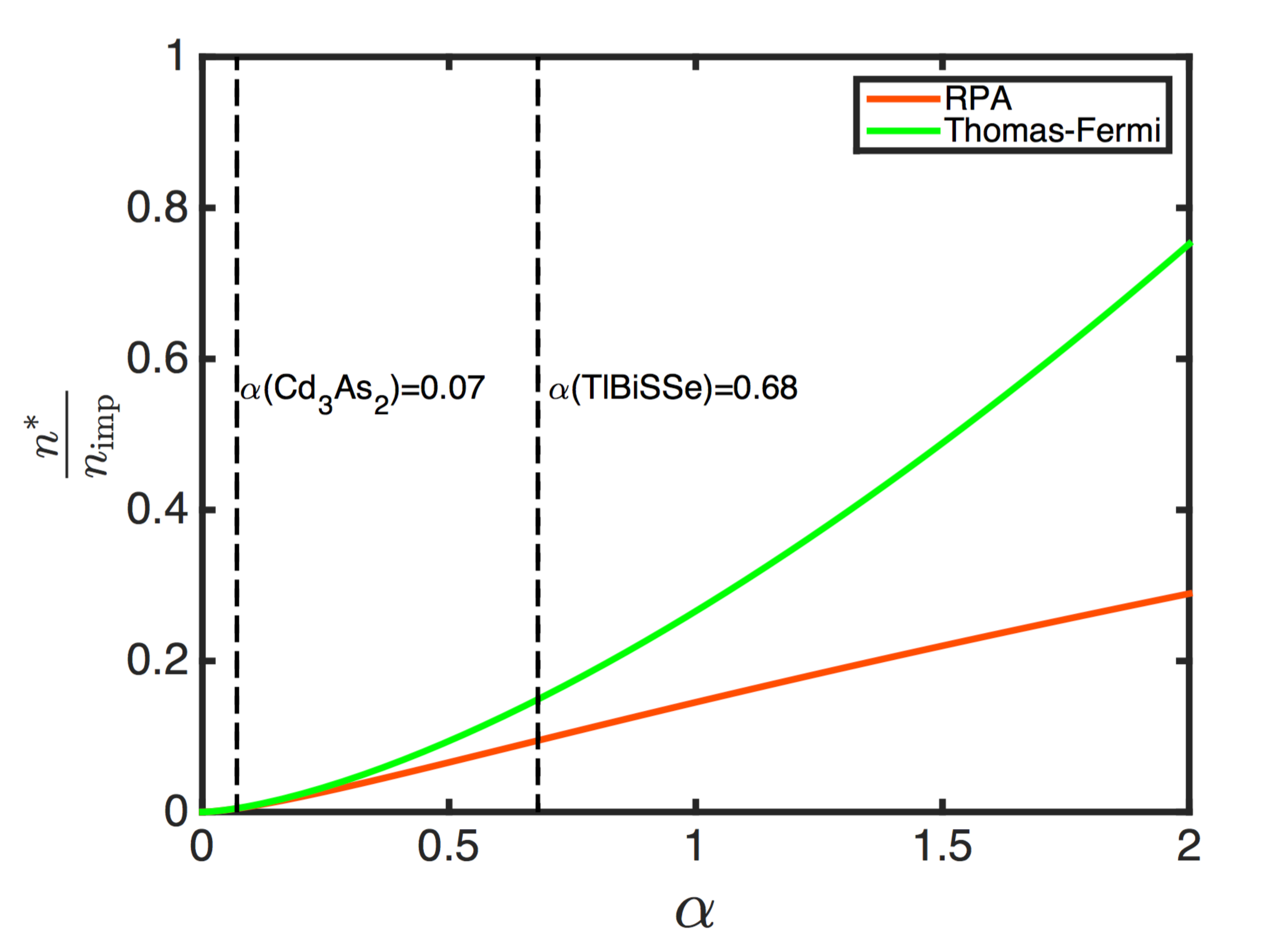}
	\caption{ {\bf Comparison of $\mathbf{n^*/n_{imp}}$ obtained using RPA and Thomas-Fermi approximations.} The error introduced by using the Thomas-Fermi screening to calculate the effective minimum carrier density, $n^*$, increases with an increase in the effective fine structure constant $\alpha$. We note that for materials such as Cd$_3$As$_2$, this is a relatively small error but this is not the case for TlBiSSe. Here we use cutoff, $\delta=10$}
	\label{fig:Fig2}
\end{figure}

The second effect that one must take into account in the inhomogeneous regime is that non-uniformly distributed impurities result in a spatially varying local chemical potential, which induces a position dependent carrier density. The induced carriers in turn screen the local potential and one eventually obtains a self consistent relationship between the local induced carrier density and the disorder averaged impurity potential, $V(\mathbf{r},n(\mathbf{r}))=\hbar v_{F}(6\pi^2 n(\mathbf{r})/g)^{1/3}$. Note that if we are far away from the Weyl point, the fluctuations in carrier density are negligible compared to the total number of carriers but close to the Weyl point this is not the case. The average induced carrier density corresponds to the average value of the random Gaussian field, $V_0=\langle V(r) \rangle_d$. 
%Finally, we note that when the average carrier density, $n_0$ and the spatial fluctuations in the carrier density $n_{rms}$ satisfy $n_0\gg n_{rms}$, or equivalently if $V_0\gg V_{rms}$, the homogeneous transport theory discussed above can be used since the fluctuations are negligible.  
%However, when $V_0 \lesssim V_{\rm rms}$, one must consider two important differences. %
%\begin{equation}% 
%\mbox{\fontsize{10}{12}\selectfont\( %
%\frac{\hbar}{\tau}=2\pi n_{\rm imp}\int\frac{d^3\mathbf{k'}}{(2\pi)^3}|U(\mathbf{q})|^2 \frac{1-\cos^2\theta}{2}\delta(E_{\rm k}-E_{\rm k'})S(\mathbf{q}, r_0). %
%\)} %
%\end{equation}
The net effect of charge doping is the appearance of macroscopic regions of charge puddles, each having an excess or a deficit of charge with respect to the average value. Therefore, one performs an averaging over these spatial fluctuations in carrier density using an effective medium theory (EMT). The EMT is a well established technique developed by Bruggerman~\citep{bruggeman1935berechnung} and later by Landauer~\cite{landauer1978electrical} in order to characterize the effects of macroscopic fluctuations on the global conductivity. In order to perform the average, one considers each macroscopic region of definite local conductivity to be embedded in a homogenous effective medium, whose conductivity is determined in a self consistent way over all the regions. Early EMT models assumed two types of regions with conductivities $\sigma_A$ and $\sigma_B$ occupying area fractions $p$ and $1-p$. This model was later generalized in the case of 2D materials, to continuous distributions of local, tensor conductivities~\citep{rossi2009effective, MoreRecentStroud}. Here we generalize the results of Ref.~\citep{MoreRecentStroud} to the continuous three dimensional case and obtain
\begin{equation} \label{eq:emt}
\int \mathscr{D} V \, P[V, V_0, V_{\rm rms}] \frac{(\hat{\sigma}(V) - \hat{\sigma}^{E})}{\left(\mathbb{\hat{I}}_3 + \frac{\mathbb{\hat{I}}_3}{3\hat{\sigma}_{xx}^{E}}(
\hat{\sigma}(V) - \hat{\sigma}^{E})\right)} = 0
\end{equation}
\begin{equation}
\hat{\sigma}=\begin{pmatrix}
\sigma_{xx}& \sigma_{xy}\\
-\sigma_{xy}& \sigma_{xx}
\end{pmatrix} \, ,\ \,
\hat{\sigma}^{E}=\begin{pmatrix}
\sigma_{xx}^{E}& \sigma_{xy}^{E}\\
-\sigma_{xy}^{E}& \sigma_{xx}^{E},
\end{pmatrix}.
\end{equation}
where $ \mathscr{D} V$ is a functional measure, $\hat{\sigma}$ is the local conductivity and $\hat{\sigma}^E$ is the effective medium conductivity to be found self consistently. In obtaining Eq.~\eqref{eq:emt} we have assumed an isotropic material i.e. $\sigma_{xx}^{E}=\sigma_{yy}^{E}=\sigma_{zz}^{E}$ and that the local conductivity regions are spherical in shape \citep{MoreRecentStroud}. This assumption is valid in the case of puddles that are small compared to the sample size. The probability distribution in Eq.~\eqref{eq:emt}, $P[V, V_0, V_{\rm rms}]$, is the same one that has been used to evaluate the Drude transport time. Therefore, it is characterized by the average and the variance of the disorder distribution (see Appendix \ref{sec:gaussian} ). Finally, we define the carrier density associated with the variance $V_{\rm rms}$ to be $n^*$ which is the effective minimum carrier density in a Weyl semimetal. In Fig.~\ref{fig:Fig2}, we show that only for $\alpha\ll 1$ does the $n^*$ calculated within the RPA reduce to that of the Thomas-Fermi. In general, as in the case of transport, the full RPA polarizability function must be used.

\subsection{Crossover Between Homogenous and Inhomogeneous Regimes}  
\label{Crossover Between Homogenous and Inhomogeneous Regimes}
In Fig.~\ref{fig:sig_vs_n}, we show the conductivity at zero magnetic field as a function of carrier density. %using $\alpha=1.2$, $g=2$ and $n_{\rm imp} = 10^{24}m^{-3}$. In this plot we choose the correlation length $r_0=0$ in order to compare our findings with those of Ref.~\citep{hwang2014carrier}. 
We find that for large carrier density, our results reproduce those of Ref.~\cite{hwang2014carrier} but at zero carrier density, and in the inhomogeneous regime, our results differ from both the conductivity calculated in Ref.~\cite{skinner2014coulomb} and Ref.~\cite{hwang2014carrier}. This is a consequence of using the EMT which provides a smooth crossover between the two regimes instead of a hard floor as done in Ref.~\cite{skinner2014coulomb}. Note that Ref.~\cite{hwang2014carrier} ignores the effect of inhomogeneity altogether. The transport theory for Weyl fermions shown in Fig.~\ref{fig:sig_vs_n} represents the full crossover from the inhomogeneous regime (close to the Weyl) point to the homogeneous regime (far away from the Weyl point).          

\section{Comparison with experiments}
\label{Comparison with experiments}

\begin{figure}[!htb]
	\centering
		\includegraphics[angle=0,width=0.50\textwidth]{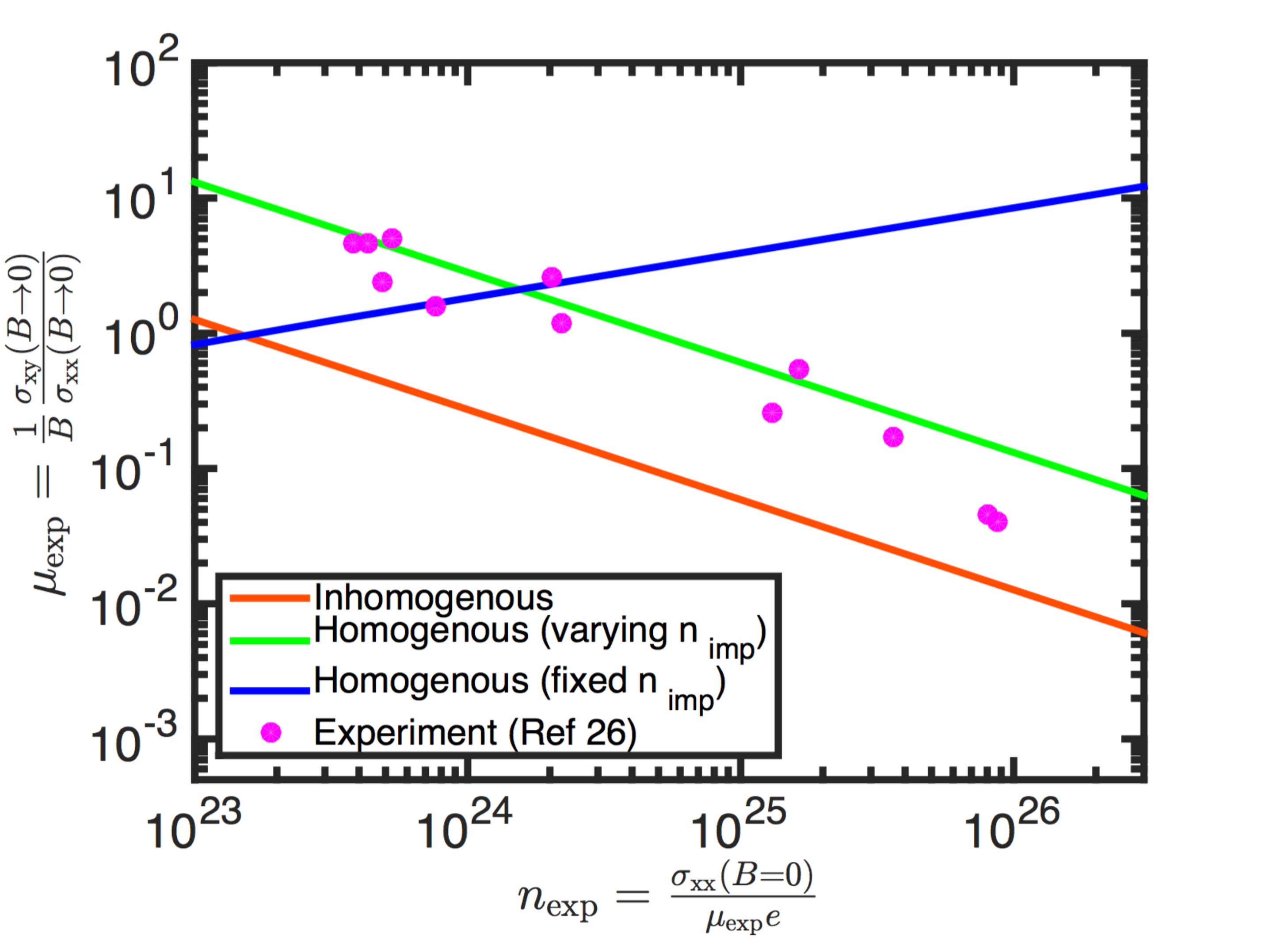}
	\caption{ {\bf Comparison with the experimental data of Ref.~\citep{novak2014large}. }%of magnetotransport measurements in TlBiSSe.  
The experimentally determined mobility of TlBiSSe, $\mu_{\rm exp}$, decreases as a function of the measured carrier density $n_{\rm exp}$, whereas the Drude theory with constant $n_{imp}$ (blue curve) would predict the opposite trend.  In the main text, we propose two possible scenarios compatible with this behavior. The red curve assumes that the experiments are in the inhomogeneous regime with $n_{\rm exp}\ll n_{\rm rms}$ with correlated charged impurities.  Alternatively, the green curve assumes that the experiments are in the homogeneous regime $n_{\rm exp}\gg n_{\rm rms}$, with the charged impurities also responsible for doping. Here $n_{\rm exp} = n_0 \approx 4~n_{\rm imp}$, $\alpha=0.68$ and $\delta=10$.}
	\label{fig:Fig3}
\end{figure}
Next, we apply our theory to address recent experimental findings on TlBiSSe ($g=4, \alpha=0.68$). Ref.~\cite{novak2014large} describe their results of a large increase in mobility with decreasing carrier density as surprising.  In order to make connection with the experimental data, one identifies the experimental mobility $\mu_{\rm exp}$ and carrier density $n_{\rm exp}$ from experimentally determined parameters~\cite{AndoPrivate}
\begin{equation}
\label{eq:nexp}
\mu_{\rm exp}=\lim_{B\to 0}\frac{\sigma_{\rm xy}(B)}{B\sigma_{\rm xx}(B)}  \, , \,  
n_{\rm exp}=\frac{\sigma_{\rm xx}(B=0)}{\mu_{\rm exp}e}.
\end{equation}
The implicit assumption made here is that the impurity concentration does not change with the sample. Indeed, for fixed $n_{\rm imp}$, Fig.~\ref{fig:sig_vs_n} shows that mobility increases with increasing carrier density. This is represented in Fig.~\ref{fig:Fig3} as the blue curve which shows the opposite trend to the experiment. We propose two possible scenarios that allow us to relax the constant $n_{\rm imp}$ in a physically justifiable way (and as we discuss below, it is not possible to determine from this data alone which of these two scenarios correspond to the experimental situation).

In the first case, we consider the experiment to be in the homogeneous regime with charged impurities also acting as dopants that shift the chemical potential \citep{burkov2011topological}. %In other words, $V_0$, and hence $n_0$, are now determined by $n_{\rm imp}$. 
The average induced carrier density is then given by $V_0=\hbar v_{F}(6\pi^2n_0/g)^{1/3} = n_{\rm imp}U(\mathbf{q}=0, n^*)$, from which we obtain $n_0 = 4~n_{\rm imp}$ in the density regime of interest. The green curve in Fig.~\ref{fig:Fig3} uses $n_{\rm exp} = n_0 = 4~n_{\rm imp}$ and shows good agreement with the experimental data. Therefore, a plausible resolution of this experimental ``mystery" is that the charged impurities that are responsible for scattering carriers are also responsible for doping the samples.  
%
%This equality holds if all the impurities are uncompensated, but in general for both positive and negative charged impurities, we can have $n_0 \leq 3 n_{\rm imp}$, although the linear scaling law should still hold.
%

A second possibility is that the samples are in the inhomogeneous regime where $n_0 \ll n_{\rm rms}$. In this regime, one equates the fluctuations in the impurity potential $V_{\rm rms}$ with the band energy to obtain an effective minimum carrier density $n^{*}$. The red curve in Fig.~\ref{fig:Fig3} considers samples with disorder ($n_{\rm imp}$ varying from $6.1\times 10^{23}m^{-3}$ to around $1.9\times 10^{27}m^{-3}$). Using the EMT equations~\eqref{eq:emt} and the definition of $n_{\rm exp}$~\eqref{eq:nexp}, we see good agreement with the experimental data. Note that here we must consider impurity position correlations and take the correlation length $r_0 \simeq L \sim n_{\rm imp}^{-1/3}$ as discussed earlier. Both the homogeneous and inhomogeneous scenarios are generally consistent with the scaling law, $\mu_{\rm exp}\sim n_{\rm exp}^{-2/3}$. This scaling is a consequence of $\sigma\sim n^{4/3}/n_{imp}$ and $n\sim n_{imp}$. Note that this also presents convincing evidence for using Coulomb impurities as opposed to neutral or point-like impurities since the conductivity scales differently with $n$ for the latter~\cite{hwang2014carrier}. We remark that our analysis is completely free of fit parameters and the both the homogeneous and inhomogeneous theories are in quantitative agreement with the data (the inhomogeneous curve agrees with the data to within a factor of 4). This also demonstrates that transport measurements alone cannot distinguish between the two regimes since both show comparable mobilities (same order of magnitude). We propose that magnetoresistance is the appropriate measurement to distinguish between the two regimes.  

\section{Magnetoresistance}  
\label{Magnetoresistance}
To calculate the magnetotransport of Weyl semimetals, we assume that the charge and the axial current are weakly coupled, since Coulomb impurities suppress the scattering between the two Weyl nodes~\cite{burkov2015negative}. Indeed, the results of Ref.~\cite{novak2014large} show no negative magnetoresistance or any effect of an in plane magnetic field, both of which are signatures of the chiral anomaly~\cite{son2013chiral,goswami2015axial}. Moreover, we note that in the absence of parallel electric and magnetic fields, the effects of the chiral anomaly are absent. Thus, the origin of transverse magnetoresistance in this system is likely due to be disorder-induced which can be treated in a semiclassical regime. We then assume that one can define a conductivity matrix with
\begin{equation}
\sigma_{xx}=\sigma_{B}({\bf r}) \frac{1}{1+\mu^2 B^2} \, ,\,
\sigma_{xy}=\sigma_{B}({\bf r}) \frac{\mu B}{1+\mu^2 B^2}, 
\end{equation}
where $\sigma_{B}({\bf r}) = \sigma(n({\bf r}))$ is the RPA-Boltzmann conductivity discussed earlier and the magnetic field is taken along the $z$ axis. 
\begin{figure}[!t]
	\centering
		\includegraphics[angle=0,width=0.45\textwidth]{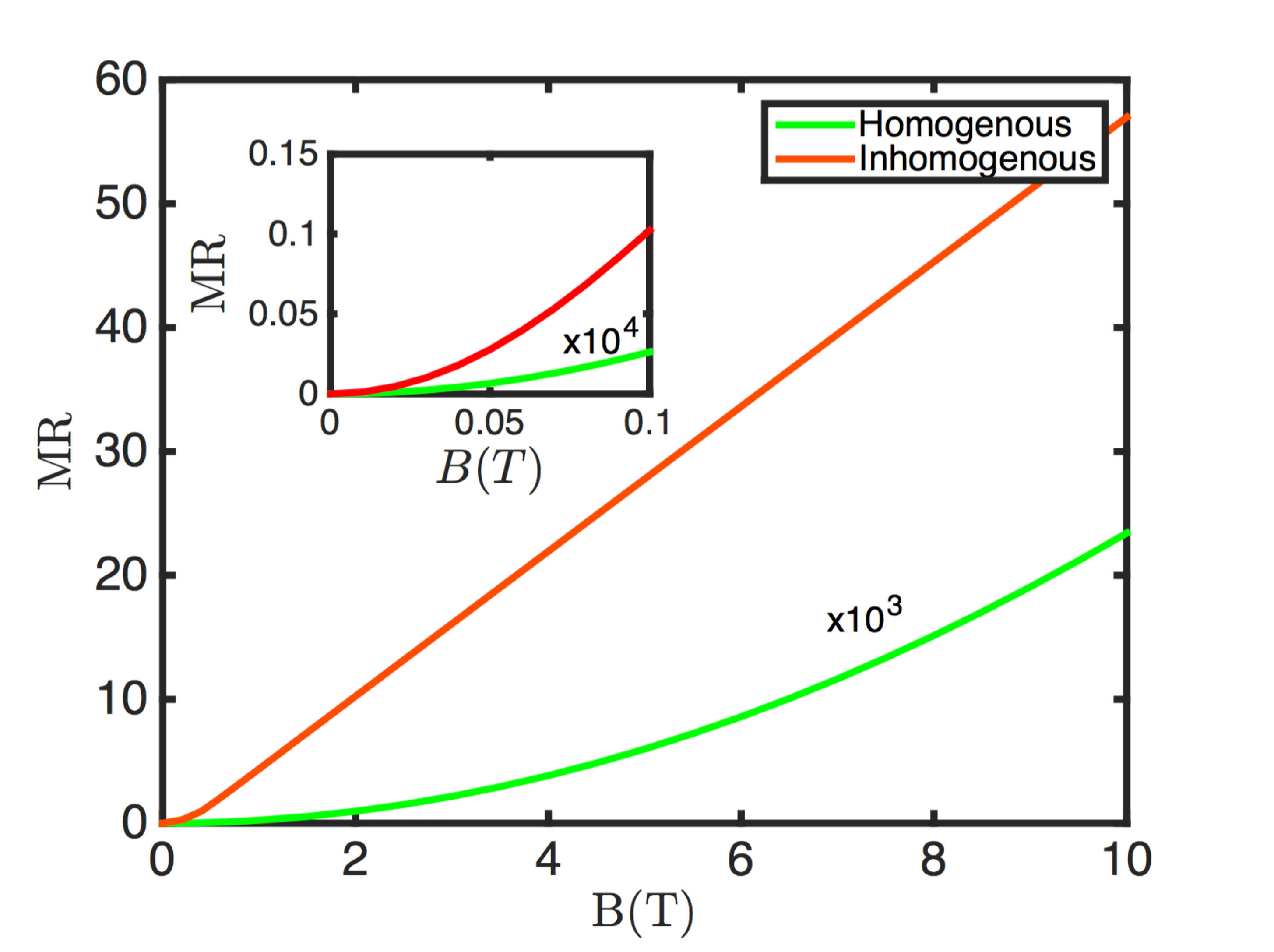}
	\caption{{\bf Magnetoresistance in TlBiSSe.} Disordered 3D Weyl fermions can have large MR $> 10$ for realistic experimental parameters.  The figure shows that both in the homogeneous and in the inhomogeneous regime, the magnetoresistance is quadratic at low fields (see inset) and linear at high fields in agreement with experimental observations. Notice that the MR in the inhomogeneous regime is much larger than that of the homogeneous regime, suggesting that increasing disorder is an easy way to enhance the MR.}
	\label{fig:Fig4}
\end{figure}
The above set of equations constitute the input for the EMT model.  The solution of the EMT equations contains both the magnetoresistance caused by having two types of carriers (electrons and holes)~\cite{hwang2007transport} and the disorder-induced magnetoresistance discussed in the context of  silver chalcogenides~\cite{parish2003non, MoreRecentStroud} and other two dimensional systems~\cite{ping2014disorder}, where $MR\equiv(\rho_{xx}(B)-\rho_{xx}(B=0))/\rho_{xx}(B=0)$. 

A simple physical picture for this disorder induced MR was proposed in Ref.~\cite{ping2014disorder} and this still holds in Weyl semimetals. Essentially, disorder results in a situation where the charge carriers move with varying drift velocities in various regions. The global Hall field thus cannot cancel the velocity dependent Lorentz force (as would be the case in a homogenous system) and the electron trajectories become longer as the magnetic field is increased. This is the origin of disorder induced magnetoresistance. 

For Weyl semimetals in general, our theory predicts that the MR should be quadratic at low magnetic fields and linear at high magnetic fields. In Fig.~\ref{fig:Fig4} we show our results for $n_0=3.8\times 10^{23}m^{-3}$ and $n_{\rm imp}=9.5\times 10^{22} m^{-3}$ (homogenous regime), and for $n_0 \rightarrow 0$,   $n_{\rm imp}=2.3\times10^{24} m^{-3}$ (inhomogeneous regime).  These values were chosen so that they correspond to similar $\mu_{\rm exp}$ and $n_{\rm exp}$ and therefore they cannot be distinguished from transport measurements alone.  The results in Fig.~\ref{fig:Fig4} demonstrate convincingly that within the semi-classical theory presented here, the magnetoresistance in the inhomogeneous regime is much larger than that of the homogeneous regime.  We also note that in the inhomogeneous regime MR is comparable to that seen by Ref.~\cite{novak2014large} for matching parameters, suggesting that those samples were in the inhomogeneous regime. Within this model, having $MR>10$ is easily achievable in the inhomogeneous regime for moderate values of B, but it is several orders of magnitude weaker in the homogeneous regime.  This suggests a clear way to experimentally distinguish between these two regimes. Moreover, an easy way to increase MR for technological applications is to make the sample dirtier, a counterintuitive, yet easily achievable, experimental goal. \\
%
%To conclude, we have developed a theory for the magnetotransport in 3D Weyl semimetals both close to the Weyl point and away from the Weyl point.  We show that recent experimental observations of decreasing $\mu_{\rm exp}$ with $n_{\rm exp}$ emerge naturally in our theory.  For reasonable experimental parameters, we predict a large classical magnetoresistance for the inhomogeneous regime that is quadratic at low magnetic fields and linear at high fields. \\

\section*{ACKNOWLEDGEMENTS} This work was supported by the National Research Foundation of Singapore under its Fellowship program (NRF-NRFF2012-01).   We would like to thank B. Feldman and S. Das Sarma for suggesting this problem, and A. Yazdani and Y. Ando for correspondence regarding their experimental data. 

\appendix
\section{Correlated Impurities}\label{sec:corr}
\begin{figure}[t!]
\center 
\includegraphics[width=0.51 \textwidth]{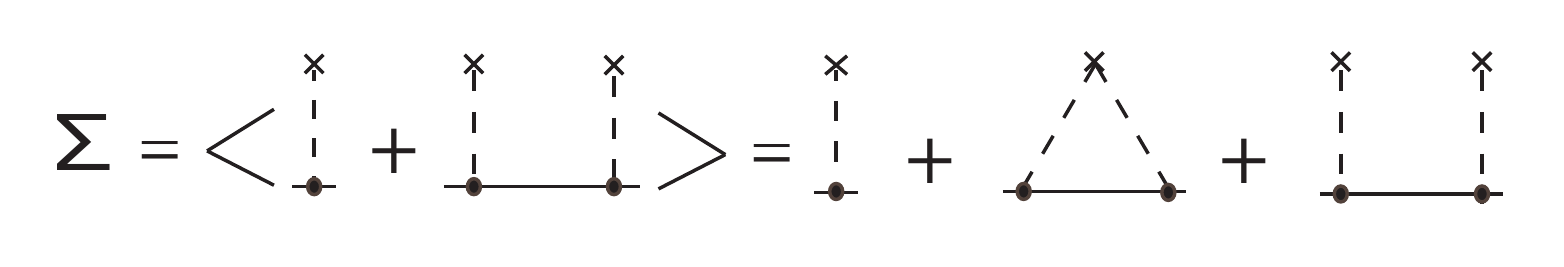}
\caption{{\bf Disorder averaged self energy}. Amputated diagrams defining the self energy at the Gaussian level. The brackets represent disorder averaging, the ``x'' accounts for the impurity concentration, the dot for the impurity potential insertion and the solid line is the 
clean propagator.}
\label{fig:self}
\end{figure}

Here we consider the effect of spatial correlations between impurities in the inhomogeneous regime. This problem has been considered before in Refs.~\cite{kawamura, li2012effect}, but its relation to diagrammatic perturbation theory is now made explicit. Disordered averaged Green's functions are defined in terms of their self energy. At the Gaussian level, the disorder averaged self energy is defined in terms of the diagrams depicted in Fig.~\ref{fig:self}. Disorder averaging is generally defined in terms of a weighted sum over impurities position of a disorder dependent quantity $O(\mathbf{R}_i)$~\cite{Rammer} 
\begin{align}\label{eq:average}
\langle O(\mathbf{r}) \rangle &= \int \prod_{i=1}^{N_{imp}} d\mathbf{R}_i \, g(\mathbf{R}_i) \, O(\mathbf{r}, \mathbf{R}_i) \\ 
\langle O(\mathbf{r})\,  O(\mathbf{r}') \rangle &= \int \prod_{i=1}^{N_{imp}} d\mathbf{R}_i  \prod_{j=1}^{N_{imp}} d\mathbf{R}_j \, 
g(\mathbf{R}_i, \mathbf{R}_j) \\ \nonumber
&\times  O(\mathbf{r}, \mathbf{R}_i) \, O(\mathbf{r}', \mathbf{R}_j) \\   \nonumber
...
\end{align}
Here $\langle ... \rangle$ stands for disorder average, $\mathbf{R}_i$ is the position of the impurities in a d-dimensional volume $L^d$, $N_{imp}$ is the total number of impurities in the system and the indices $i,j,l...$ label different impurities. The important objects in the above definitions are the correlation functions $g(\mathbf{R}_i, \mathbf{R}_j, ..., \mathbf{R}_z)$, describing correlations between one impurity, two impurities and so on \cite{ziman1979models}. As usual, the hierarchy of correlation functions cannot be worked out explicitly and one has to perform some physically 
motivated ansatz in order to truncate the hierarchy. For completely random disorder, all correlation functions factorize as product of 
single particle correlations $g(\mathbf{R}_i)$~\cite{Rammer}. These are simply equal to the probability of finding an impurity at site $i$, i.e. $1/L^d$. The pair correlation function $g(\mathbf{R}_i, \mathbf{R}_j)$ gives the probability of finding an impurity at site $R_i$ given one at site $R_j$.    
Let us consider the total disorder potential $V(\mathbf{r})$ defined in Eq.~\eqref{Ham} 
\begin{equation}\label{eq:potimp}
V(\mathbf{r}) = \sum_{i=1}^{N_{imp}} \, U(\mathbf{r}-\mathbf{R}_i).
\end{equation}
In the weak scattering limit (Born limit), only the first two moments of the distribution of  $V(\mathbf{r})$ are relevant, corresponding to the diagrams of Fig.~\ref{fig:self}. Performing the impurity average, one obtains three diagrams: two proportional to $N_{imp}$ and one to $N_{imp}^2$. For completely random disorder, the term proportional to $N_{imp}^2$ is a constant that can be renormalized away.  However, as we are now going to show, if there is any residual structure in the disorder, the $N_{imp}^2$ term cannot be renormalized away.\\

For completeness, we start considering the standard single impurity diagram; in momentum space it reads:
\begin{equation} \label{eq:first} 
\sum_{i=1}^{N_{imp}} \langle U(\mathbf{k}-\mathbf{k}') e^{-\imath \, \mathbf{R}_i (\mathbf{k}-\mathbf{k}')} \rangle = n_{imp} \, U(0) \, \delta(\mathbf{k}-\mathbf{k}'),
\end{equation}
where $n_{imp}=N_{imp}/L^d$ is the density of impurities and $U(0)$ is the impurity potential evaluated at zero transferred momentum $q$. Note that in principle this term is singular and needs to be regularized, for example by screening. Here we will assume that the impurity potential is screened and therefore consider $U(0)$ as a finite quantity. Consider now the second order term
\begin{align} \label{eq:second} \nonumber
\sum_{i,j=1}^{N_{imp}} \frac{1}{L^d}\sum_{k'} &\langle U(\mathbf{k}-\mathbf{k}') \, e^{-\imath \, \mathbf{R}_i (\mathbf{k}-\mathbf{k}')} \, G_0(\mathbf{k}',E) \\ 
&\times \, U(\mathbf{k}'-\mathbf{k}'')\, e^{-\imath \, \mathbf{R}_j (\mathbf{k}'-\mathbf{k}'')} \rangle,
\end{align}
where $G_0(\mathbf{k}',E)$ is the free electron propagator. Taking the disorder average, there are two contributions: for $i=j$ one finds
\begin{equation}
 \sum_{i=1}^{N_{imp}}  \langle  e^{-\imath \, \mathbf{R}_i (\mathbf{k}-\mathbf{k}')} \, e^{-\imath \, \mathbf{R}_j (\mathbf{k}'-\mathbf{k}'')} \rangle = n_{imp} \, \delta(\mathbf{k}-\mathbf{k}''), 
\end{equation}
corresponding to the rainbow diagram shown in Fig.~\ref{fig:self}. For $i \neq j$, we need to make some assumptions on the pair correlation function. We assume that it still depends only on the coordinate difference $\mathbf{R}_i-\mathbf{R}_j$~\cite{ziman1979models}. As a consequence, momentum is conserved on average, i.e. $\mathbf{k}= \mathbf{k}''$. We also assume that the pair correlation function does not depend on the angle between $\mathbf{R}_i$ and $\mathbf{R}_j$. Within these assumptions, Eq.~\eqref{eq:second} reads
\begin{align} \label{eq:secondF}
\frac{1}{L^d} \sum_{\mathbf{k}'} G_0(\mathbf{k}', & E) \, U^2(|\mathbf{k}-\mathbf{k}'|) \, n_{imp} \\ \nonumber
&\times \left\{ 1+ n_{imp} \, \int d^d R \, g(R) e^{-\imath \, \mathbf{R} (\mathbf{k}-\mathbf{k}')} \right\},
\end{align}
where the first term in the curly bracket corresponds to the Born term (single particle scattering) and the second takes into account the effect of two particles scattering. Note that the second term is singular at $\mathbf{k}=\mathbf{k}'$ \cite{ziman1979models}; to take care of this singularity, one subtracts it as the Fourier transform of unity and we can define the regularised structure factor as
\begin{equation}\label{eq:structure}
S(|\mathbf{k}-\mathbf{k}'|) = 1+ n_{imp} \, \int d^d R \, [ g(R)-1] \, e^{-\imath \, \mathbf{R} (\mathbf{k}-\mathbf{k}')}.
\end{equation}
This has been obtained in Refs.~\cite{kawamura, li2012effect} and the structure factor can be measured e.g. in neutron diffraction experiments. Next, we consider the case where the term proportional to $n_{imp}^2$ is important. For the charge carriers, the ``size" of the charged impurity is roughly given by the effective Bohr radius $a_0$ of its lowest impurity level; this can be orders of magnitude larger~\footnote{An impurity concentration of $10^{-4}-10^{-2}$ impurities per unit volume should be considered quite dense according to this principle.} than the underlying lattice constant~\cite{ziman1979models}. If one compares the average spacing $L$ between the impurities and $a_0$, one comes at the conclusion that correlation effects are important if $L<a_0$. Note that the ratio $L/a_0$ is reminiscent of the parameter $r_s$ used to quantify correlations in an electron gas. We conclude by connecting the above analysis to the relaxation time $\tau$. By definition $1/\tau(k)={\rm Im} \, \Sigma(k)$, where 
\begin{equation}\label{eq:rel1}
{\rm Im} \, \Sigma(k)= \frac{n_{imp}}{L^d} \sum_{\mathbf{k}'}  \, U^2(|\mathbf{k}-\mathbf{k}'|)  \, S(|\mathbf{k}-\mathbf{k}'|) \, {\rm Im} G_0(\mathbf{k}',E)
\end{equation}
It follows that the variance of the random Gaussian field $V(\mathbf{x})$ can be written as 
\begin{equation}\label{eq:rel2}
\langle V(\mathbf{r}) V(\mathbf{r}') \rangle_d = n_{imp} \int d^3q \, e^{\imath \, \mathbf{q} (\mathbf{r}-\mathbf{r}')} \, U^2(q) \, S(q)
\end{equation}
that is our Eq.~\eqref{eq:rel} of the main text.

\section{On the Gaussian approximation} \label{sec:gaussian}
Here we consider the form of the disorder probability distribution used in Eq.~\eqref{eq:emt} of the main text. As explained in the main text, within the Drude transport theory, this function is the same as the one used to evaluate the transport time. Here we provide an explicit proof of the validity of the Gaussian approximation. This discussion is mostly based on Ref.~\cite{galitski2007statistics} and is based on the functional approach to disordered systems. This method is completely equivalent to the ``sum over impurities" approach used in diagrammatic perturbation theory. In the functional approach, one is interested in evaluating the disordered averaged generating functional  
\begin{equation} \label{eq:def1}
\langle \log Z[V] \rangle_d = \int \mathscr{D} V \, P[V] \, \log Z[V],
\end{equation}
where $\mathscr{D} V$ is a functional measure. Here we are not interested in the actual calculation of this quantity, but only in finding $P[V]$. Generally, $\log Z[V]$ can be substituted with an effective functional of $V$ as in the case of the Effective Medium Theory (EMT) employed in the main text. Let us consider the total disorder potential  of Eq.~\eqref{eq:potimp}, where the potential $U$ is due to the $N_{imp}$ impurity potentials and $V(\mathbf{r})$ is the resulting effective potential. In order to find $P[V]$, one needs to interpret the above definition as a constraint. In the functional formalism this is accomplished by means of a functional Dirac delta function averaged over impurity positions
\begin{align}\label{eq:functV}
P[V] &= \prod_{\mathbf{r}} \langle \delta [ V(\mathbf{r}) - \sum\limits_{i=1}^{N_{imp}} \, U(\mathbf{r}-\mathbf{R}_i) ] \rangle \\ \nonumber
&= \int \mathscr{D} \xi \, \bigl \langle  e^{\imath \int d^3r \, \xi(\mathbf{r}) [ V(\mathbf{r}) - \sum\limits_{i=1}^{N_{imp}} \, U(\mathbf{r}-\mathbf{R}_i) \, ]} \, \bigr \rangle,
\end{align} 
where we have used the functional representation of the Dirac delta function and $\xi(\mathbf{r})$ is a Lagrange multiplier field. The first step in the evaluation of $P[V]$ consists in assessing the correlated nature of the impurities. According to Ref.~\cite{Rammer}, and as explained in the main text, if the distance between the impurities is much larger than the screening length, then the positions of the impurities are uncorrelated. On the other hand, if the screening length of the Coulomb potential is comparable to the average distance between the impurities, then there may be correlations in the positions of the impurities themselves (see Appendix~\ref{sec:corr}). Here we consider for simplicity the case of completely uncorrelated disorder and comment on the effect of correlations at the end. In the thermodynamic limit one obtains~\cite{galitski2007statistics} 
\begin{widetext}
\begin{equation}
P[V] =  \int \mathscr{D}\xi \, e^{i \, \int d^3r \, \xi(\mathbf{r}) \, V(\mathbf{r})} \exp \left\{- n_{imp} \int d^3 R 
\left(1- e^{-\imath \int d^3 r \, \xi(\mathbf{r}) \, U(\mathbf{r}-\mathbf{R})} \right) \right\} = \int  \mathscr{D}\xi \, e^{i \, \int d^3 r \,\xi(\mathbf{r}) \, V(\mathbf{r})} \, e^{\Phi(\xi)}, 
\end{equation}
\end{widetext}
where in the second step we have identified the cumulant function of the stochastic process $\Phi(\xi)$ and the related characteristic function $\chi(\xi)=e^{\Phi(\xi)}$. We now follow Ref.~\cite{kubo} (where more details can be found) and assess the Gaussian nature of the characteristic function using the central limit theorem. One needs to expand the cumulant function and show that the magnitude of the second cumulant is the dominant one. This is basically the well known Born criterion, c.f. Ref.~\cite{Rammer}. For screened Coulomb disorder, and for the considered values of the electromagnetic coupling $\alpha$, the Born criterion is satisfied for carrier density $n \geq n^*$, see Section~\ref{sec:emt}. Finally, integrating out the Lagrange multiplier field, one finds the Gaussian distribution of the random fields used in the main text. When $n \simeq n^*$, the screening length of the Coulomb potential is comparable to the average distance between the impurities and one needs to use the modified propagator of the random fields, Eq.~\eqref{eq:rel2}.
\bibliography{weylbib4}

\begin{thebibliography}{49}
\bibitem [{cut()}]{cutoff}%
  \BibitemOpen
  \href@noop {} {\bibinfo  {journal} {In this work, the results of Fig. 2 and
  Fig. 3 are obtained using $\delta=10$ following Ref. 20}\ }\BibitemShut
  {NoStop}%
\end{thebibliography}

\end{document}